\documentclass[12pt]{article}
\usepackage[reqno]{amsmath}
\usepackage{epsfig}
\usepackage{array}
\usepackage{float}
\usepackage{lscape,graphicx} 
\usepackage{graphics}
\usepackage{amssymb}
\usepackage{color}
\usepackage{multirow}

\def\gtap{\ \raisebox{-.4ex}{\rlap{$\sim$}} \raisebox{.4ex}{$>$}\ }

\newcommand{\bea}{\begin{equation}\begin{array}{c}}
\newcommand{\eea}{\end{array}\end{equation}}
\newcommand{\ea}{\end{array}}

\newcommand{\beq}{\begin{equation}}
\newcommand{\eeq}{\end{equation}}
\newcommand{\bad}{\begin{array}{ccc}}

\newcommand{\ba}{\begin{array}{c}}
\newcommand{\half}{\frac{1}{2}}
\newcommand{\diag}{{\rm diag}}

\begin{document}
\noindent \vspace{1cm}
\hfill{{\small FLAVOUR(267104)-ERC-23}}
\hfill{{\small TUM-HEP 850/12}} 
\hfill{{\small SISSA 25/2012/EP}}
\hfill{{\small CFTP/12-013}}
%
%\hfill{{\small Ref. CFTP/12-007}}    

%\rightline{{\small arXiv:12ZZ.xxxx[hep-ph]}}

\begin{center}
\mathversion{bold}
{\bf{\Large Higgs Decays in the Low Scale 
\\ Type I See-Saw Model}}
\mathversion{normal}

\vspace{0.4cm}
C. Garcia Cely$\mbox{}^{a)}$,
A. Ibarra$\mbox{}^{a)}$,
E. Molinaro$\mbox{}^{b)}$ and
S. T. Petcov$\mbox{}^{c,d)}$~\footnote{Also at: 
Institute of Nuclear Research and
Nuclear Energy, Bulgarian Academy of Sciences, 1784 Sofia, Bulgaria}

\vspace{0.2cm}
$\mbox{}^{a)}${\em Physik-Department T30d, Technische 
Universit\"at M\"unchen,\\ 
James-Franck-Stra{\ss}e, 85748 Garching, Germany.\\}

\vspace{0.1cm}
$\mbox{}^{b)}${\em  Centro de F\'{i}sica Te\'{o}rica de Part\'{i}culas, \\
Instituto Superior T\'{e}cnico, Technical University of Lisbon, \\ 1049-001,
Lisboa, Portugal. \\}

\vspace{0.1cm}
$\mbox{}^{c)}${\em  SISSA and INFN-Sezione di Trieste, \\
Via Bonomea 265, 34136 Trieste, Italy.\\}

\vspace{0.1cm}
$\mbox{}^{d)}${\em Kavli IPMU, University of Tokyo (WPI), Tokyo, Japan.\\}

\end{center}

\begin{abstract}
The couplings of the low scale type I see-saw model are severely
constrained by the requirement of reproducing the correct neutrino
mass and mixing parameters, 
by the non-observation of lepton number and charged 
lepton flavour violating processes and by electroweak precision data.
We show that all these constraints still
allow for the possibility of an exotic Higgs decay channel into a light 
neutrino and a heavy neutrino with a sizable branching ratio. 
We also estimate the prospects to observe this decay 
at the LHC and discuss its complementarity to the indirect
probes of the low scale type I see-saw model from experiments
searching for the $\mu\rightarrow e\gamma$ decay.
\end{abstract}

\section{Introduction}
It is well established experimentally on the basis of 
the neutrino oscillation data that neutrinos have  
non-zero masses which are much smaller than the 
charged lepton and quark masses, 
and that they mix. The enormous disparity 
between the magnitude of the neutrino masses 
and the masses of the charged leptons and quarks suggests 
that the neutrino masses are related to the existence 
of a new mass scale in physics, i.e., 
to new physics beyond the Standard Model (SM). 
 The simplest extension of the SM,
which allows to explain naturally the smallness 
of the neutrino masses and the existence 
of neutrino mixing, consists of 
introducing two right-handed (RH) 
fermions as $SU(2)_{L}\times U(1)_{Y}$ singlets,
usually known as RH neutrinos, which 
have Yukawa-type couplings with the SM Higgs 
and left-handed (LH) lepton doublets. 
Unless one imposes additional ad-hoc (global) symmetries, 
the RH neutrinos have also a Majorana mass term which 
breaks explicitly the total lepton charge  conservation.
In such type I see-saw scenarios \cite{seesaw}, 
the light neutrinos are therefore  predicted to be Majorana 
particles and their small masses are generated after the 
electroweak (EW) symmetry breaking due to the interplay  
between the neutrino Yukawa couplings and the 
Majorana masses of the RH neutrinos. 
The scale $\Lambda$ at which the new physics 
manifests itself, which is set by the scale of 
masses of the RH neutrinos, can, in principle,
have an arbitrary large value, up to the GUT scale of 
$2\times 10^{16}$ GeV and even beyond, up to 
the Planck mass. An interesting possibility, which can also 
be theoretically well motivated (see, e.g., 
{\cite{Shaposhnikov:2006nn,Kersten:2007vk,Gavela:2009cd}), 
is to have the new physics at the TeV scale, 
i.e., $\Lambda\sim(100 -1000)$ GeV. 
Low scale see-saw scenarios usually predict a 
rich phenomenology at the TeV scale
and are constrained by different sets of data,
such as, e.g., the data on neutrino oscillations, 
from EW precision tests and on the 
lepton flavour violating (LFV) processes 
$\mu \rightarrow e \gamma$, $\mu \rightarrow 3e$, 
$\mu^- - e^-$ conversion in nuclei.
In the case of the TeV scale type I see-saw 
scenario of interest, the flavour structure of the couplings of 
the heavy Majorana neutrinos $N_{1}$ and $N_2$ 
to the charged leptons and the $W^{\pm}$ bosons, and 
to the LH flavour neutrinos $\nu_{\ell L}$ and the $Z^0$ boson,
are essentially determined by the requirement of 
reproducing the data on the neutrino
oscillation parameters \cite{Ibarra:2011xn}
(see also \cite{Raidal:2004vt,Gavela:2009cd}). 
The strongest constraints on the parameter space 
of this scenario is provided by the 
data on the $\mu \rightarrow e \gamma$ and 
$\mu \rightarrow 3e$ decays and the
$\mu^- - e^-$ conversion in nuclei
\cite{Ibarra:2010xw,Ibarra:2011xn,Dinh:2012bp}. 
Given the constraints on the neutrino Yukawa couplings 
which follow from the current upper bound on the $\mu\to e\gamma$ 
decay rate \cite{MEG},
the charged current (CC) and neutral current (NC) weak 
interaction couplings of the heavy Majorana neutrinos $N_{1,2}$
are not sufficiently large to allow their direct production at 
the LHC with an observable rate \cite{Ibarra:2011xn}. 

In this Letter we consider the possibility of producing 
these new fermions from Higgs boson decays, in the scenario 
in which the see-saw mass scale is smaller than the 
Higgs boson mass.\footnote{ A similar study has been 
done recently in \cite{Dev:2012zg} in the context 
of inverse see-saw models with heavy singlet fermions at the EW scale. 
However, in the analysis performed in \cite{Dev:2012zg} 
the relevant constraints on the see-saw parameter 
space and the limits on the Yukawa couplings, which arise from
neutrino oscillation data and the experimental 
searches of charged lepton flavour violation, 
were not included. Higgs decays in RH neutrinos were also considered in \cite{Pilaftsis:1991ug,Chen:2010wn} 
in a model in which the neutrino masses are generated at one loop level.}
Current collider searches exclude at the 95\% C.L. 
Higgs masses below 114.4 GeV (LEP \cite{Barate:2003sz})
and the windows 127 GeV to 600 GeV (CMS \cite{Chatrchyan:2012tx}), 
111.4 GeV to 116.6 GeV, 119.4 GeV to 122.1 GeV, 
and 129.2 GeV to 541 GeV (ATLAS \cite{:2012an}). 
We will concentrate here on the low mass allowed window and we will take 
as benchmark value a Higgs mass
$m_{h}=125$ GeV, which is in agreement with the new particle recently
discovered by the ATLAS and CMS experiments \cite{:2012gk,:2012gu}, 
and which is at the moment a good candidate for a Standard Model Higgs boson.
In this framework, the presence of a new Higgs boson decay channel, 
with heavy Majorana neutrinos in the final state, 
does not modify the SM Higgs boson production mechanisms  at LHC, 
but enlarge the total Higgs decay width, thus lowering the 
decay branching ratios predicted in the Standard Model. 
We consider what are the constraints that one can impose
on the size of neutrino Yukawa couplings in these scenarios 
from a possible observation of the new decay channel at LHC 
 as well as the interplay with the limits obtained 
using the data from the experiments on LFV processes 
involving the charged leptons.

 The text is organized as follows: 
in section 2 we recapitulate the formalism and discuss 
the relevant parameter space in type I see-saw 
scenarios with RH neutrino masses at the electroweak scale. 
In section 3 we discuss the new Higgs decay channel and in section 4 we
analyze quantitatively the prospects for production and
detection of the heavy RH neutrinos in Higgs decays at the LHC. 
All the relevant results are summarized in the 
last section of the Letter.

\section{Preliminary Remarks}

 The light neutrino Majorana mass matrix is generated 
from the following Lagrangian, arising 
in type I see-saw extensions of the SM:
%%%%%%%%%%%%%%%%%%%%%%%%%%%%%%%%%%%%%%%%%%%
\begin{equation}
\mathcal{L}_{\nu}\;=\; -\, \overline{\nu_{\ell L}}\,(M_{D})_{\ell a}^{*}\, \nu_{aR} - 
\half\, \overline{\nu^{C}_{aL}}\,(M_{N})_{ab}^{*}\,\nu_{bR}\;+\;{\rm h.c.}\,, 
\label{typeI}
\end{equation}
%%%%%%%%%%%%%%%%%%%%%%%%%%%%
%
where $\nu^{C}_{aL}\equiv C \overline{\nu_{aR}}^T$, $C$ being the charge 
conjugation matrix, $M_{N} = (M_{N})^T$ is the 
$k\times k$ Majorana mass matrix of the right-handed 
(RH) neutrinos $\nu_{aR}$,
and $M_{D}$ is a $3\times k$ neutrino Dirac 
mass matrix which is generated by the matrix of 
neutrino Yukawa couplings after the electroweak 
(EW) symmetry breaking.  In the following we consider 
the TeV scale type I see-saw scenarios 
with two RH neutrinos discussed in~\cite{Ibarra:2010xw,Ibarra:2011xn,Dinh:2012bp}.\footnote{Type I see-saw scenarios with two 
heavy Majorana neutrinos having masses by few to several orders 
of magnitude below the GUT scale of $\sim 2\times 10^{16}$ GeV  
have been discussed, e.g., in 
\cite{3X2Models,Ibarra:2003up,PRST05}.}

Taking into account eq.~(\ref{typeI}), and working in the basis 
in which the RH neutrino mass matrix is diagonal,
the couplings of the heavy Majorana neutrino 
mass eigenstates $N_{1}$ and $N_{2}$ with the 
SM leptons and the 
SM Higgs boson $h$ are given by:
%%%%%%%%%%%%%%%%%%%%%%%%%%%%%%%%%%%%%%%%%%%
\begin{eqnarray}
\mathcal{L}_{H}^N &=& -\frac{g M_{k}}{4\, M_{W}}\,
\overline{\nu_{\ell L}}\,(RV)_{\ell k}\,(1 + \gamma_5)\,N_{k}\,h\;
+\; {\rm h.c.}\,,
\label{NH}
\end{eqnarray}
%%%%%%%%%%%%%%%%%%%%%%%%%%%%%%%%%%%%%%
%
where $R \simeq (M_D\, M^{-1}_{N})^*$ and $V$ is the unitary matrix that
diagonalises the RH neutrino mass matrix,
$M_{N} \simeq V^{*} \diag(M_{1},M_{2}) V^{\dagger}$, 
with $M_{1,2}>0$.
The combination $(RV)$ parametrises the mixing 
between the SM active left-handed (LH) 
flavour neutrinos $\nu_{\ell L }$ and 
the SM singlet RH neutrinos $\nu_{aR}$ and
determines the charged current and the neutral current 
weak interaction couplings of the heavy
Majorana neutrinos $N_k$ to the $W^\pm$ and $Z^0$ bosons:
%%%%%%%%%%%%%%%%%%%%%%%%%%%%%%%%%%%%%%%%%%%
\begin{eqnarray}
 \mathcal{L}_{CC}^N &=& -\,\frac{g}{2\,\sqrt{2}}\,
\bar{\ell}\,\gamma_{\alpha}\,(RV)_{\ell k}(1 - \gamma_5)\,N_{k}\,W^{\alpha}\;
+\; {\rm h.c.}\,
\label{NCC},\\
 \mathcal{L}_{NC}^N &=& -\frac{g}{4 \,c_{w}}\,
\overline{\nu_{\ell L}}\,\gamma_{\alpha}\,(RV)_{\ell k}\,(1 - \gamma_5)\,N_{k}\,Z^{\alpha}\;
+\; {\rm h.c.}\,
\label{NNC}
\end{eqnarray}
%%%%%%%%%%%%%%%%%%%%%%%%%%%%%%%%%%%%%%

The elements of the matrix $(RV)$  
should satisfy the following constraint
which is characteristic of the type I see-saw
mechanism under discussion 
%%%%%%%%%%%%%%%%%%%%%%%%%%%%%%%%%%%%%%%%%%%
\begin{equation}
|\sum_{k} (RV)^*_{\ell'k}\;M_k\, (RV)^{\dagger}_{k\ell}|
\simeq |(m_{\nu})_{\ell'\ell}| \lesssim 1~{\rm eV}\,,
~\ell',\ell=e,\mu,\tau\,.
\label{VR1}
\end{equation}
%%%%%%%%%%%%%%%%%%%%%%%%%%%%%%%%%%%%%%%%%%
%
Here $m_{\nu}$ is the Majorana mass matrix of the LH
flavour neutrinos generated by the see-saw mechanism.
The upper limit $|(m_{\nu})_{\ell'\ell}| \lesssim 1$ eV,
$\ell,\ell'=e,\mu,\tau$, follows from the existing data
on the neutrino masses and on the neutrino mixing
\cite{Merle:2006du}.  For the values of the masses $M_k$
of the heavy Majorana neutrinos $N_k$ of interest
for the present study, $M_k\lesssim 125$ GeV,
the simplest scheme in which the constraint
(\ref{VR1}) can be satisfied is \cite{Ibarra:2011xn}
that in which the two heavy Majorana neutrinos
$N_1$ and $N_2$ form a pseudo-Dirac neutrino $N_{PD}$
\cite{LW81,STPPD82}: $M_2 = M_1(1 + z)$, where $z \ll 1$, 
and $N_{PD} = (N_1 \pm i\,N_2)/\sqrt{2}$.
In the scenario where the CC and NC couplings of $N_{1,2}$
are ``sizable'' leading to observable effects 
at low energies, the requirement of reproducing 
the correct neutrino oscillation parameters determines the  
couplings $(RV)_{\ell 1}$ and  $(RV)_{\ell 2}$ in 
eqs. (\ref{NCC}) and  (\ref{NNC}).
The concrete expressions depend on whether the 
neutrino masses exhibit
a normal hierarchy (NH) 
or an inverted hierarchy (IH) and read~\cite{Ibarra:2011xn}:
%%%%%%%%%%%%%%%%%%%%%%%%%%%%%%%
\begin{eqnarray}
\label{mixing-vs-y}
\left|\left(RV\right)_{\ell 1} \right|^{2}&=&
\frac{1}{2}\frac{y^{2} v^{2}}{M_{1}^{2}}\frac{m_{3}}{m_{2}+m_{3}}
    \left|U_{\ell 3}+i\sqrt{m_{2}/m_{3}}U_{\ell 2} \right|^{2}\,,
~~{\rm NH}\,,\\
\left|\left(RV\right)_{\ell 1} \right|^{2}&=&
\frac{1}{2}\frac{y^{2} v^{2}}{M_{1}^{2}}\frac{m_{2}}{m_{1}+m_{2}}
    \left|U_{\ell 2}+i\sqrt{m_{1}/m_{2}}U_{\ell 1} \right|^{2}
\simeq \;\frac{1}{4}\frac{y^{2} v^{2}}{M_{1}^{2}}
\left|U_{\ell 2}+iU_{\ell 1} \right|^{2}\,,
\,{\rm IH}\,,
\label{mixing-vs-yIH}\\
(RV)_{\ell 2}&=&
\pm i\, (RV)_{\ell 1}\sqrt{\frac{M_1}{M_2}}\,,~\ell=e,\mu,\tau\,,
\label{rel0}
\end{eqnarray}
%%%%%%%%%%%%%%%%%%%%%%%%%%%%
%
where $v\simeq174$ GeV and in 
eq. (\ref{mixing-vs-yIH}) we have used 
the fact that for the IH spectrum one has $m_1 \simeq m_2$. 
The parameter $y$ in the expressions above represents the largest
eigenvalue of the matrix of neutrino Yukawa couplings  $m_D/v$
\cite{Ibarra:2011xn}:
%%%%%%%%%%%%%%%%%%%%%%%%%%%%%%%%%
\begin{equation}
\label{ymax2}
y^{2}v^{2}\;=\;2\,M_{1}^{2}\,\left(\left| (RV)_{e1} \right|^{2}+
\left| (RV)_{\mu1} \right|^{2}+\left| (RV)_{\tau1} \right|^{2}\right)\,.
\end{equation}  
%%%%%%%%%%%%%%%%%%%%%%%
%

 For $M_{1,2}\lesssim 125$ GeV, the most stringent 
upper limits on  $|(RV)^*_{e1} (RV)_{\mu1}|$, and thus on 
the magnitude of $y$, can be obtained
from the existing experimental upper bound
on the rate of the lepton flavour violating (LFV)
process $\mu\to e\gamma$ \cite{Ibarra:2011xn,Dinh:2012bp}.
Taking the best fit values of the neutrino oscillation 
parameters \cite{Tortola:2012te}, we get the upper limits:
\begin{eqnarray}
&& y\lesssim 0.042\,~{\rm for~NH~with~} M_1=100\,{\rm GeV}\,,
\label{yupNH}\\
&& y\lesssim 0.056\,~{\rm for~IH~with~} M_1=100\,{\rm GeV}\,.
\label{yupIH}
\end{eqnarray}
These upper limits are roughly of the same order as the 
bottom Yukawa coupling, $y_{b}=m_{b}/v\simeq 0.024$.
It is then interesting to explore the impact of the heavy neutrinos
with a possibly sizable Yukawa coupling in the Higgs
phenomenology. In this Letter we will discuss the possibility 
of observing the exotic Higgs decays 
$h\to \nu_{\ell L} + \overline{N_{PD}},\;\overline{\nu_{\ell L}} + N_{PD}$ 
at the LHC. For brevity we will denote these decays
generically as $h\to \nu N$ in what follows.

%%%%%%%%%%%%%%%%%%%%%%%%%%%%%%%%%%%
\begin{figure}[t]
\begin{center}
\includegraphics[width=16cm,height=10cm]{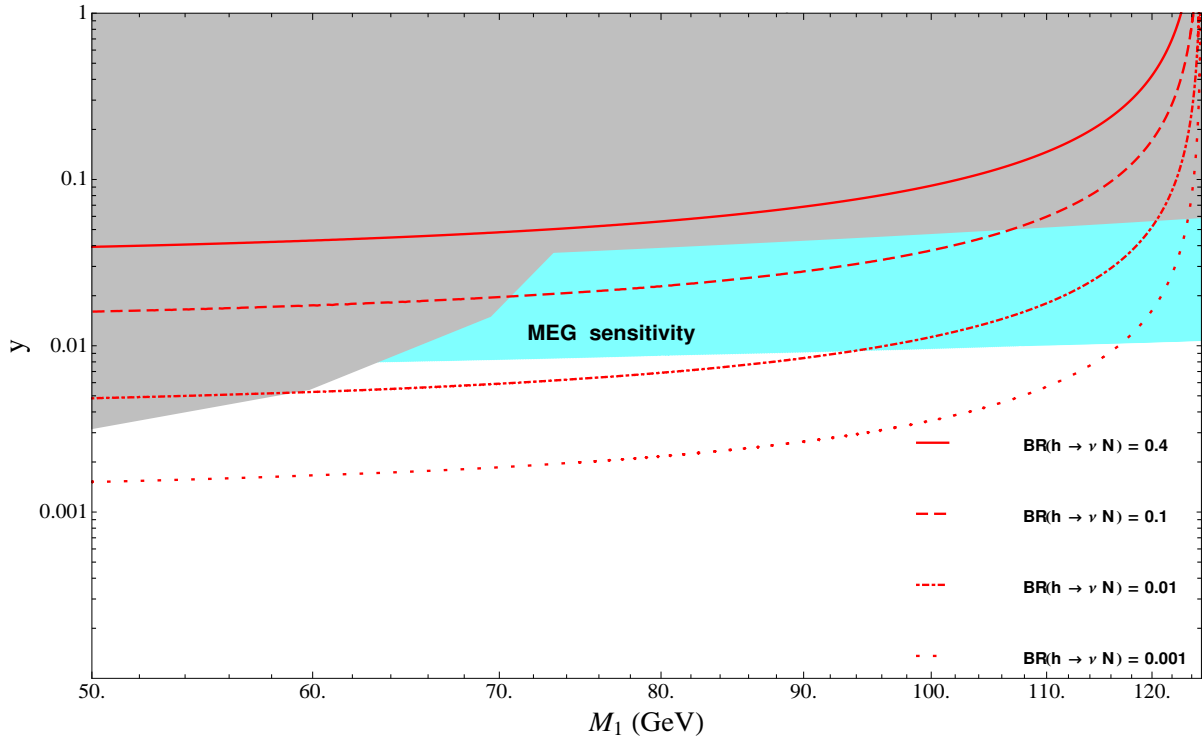}
\caption{Values of the neutrino Yukawa coupling $y$ probed 
by Higgs decays into $N_{PD}$ for $m_{h}=125$ GeV. 
The grey region is excluded by LEP2 data \cite{LEP2} and  
searches of lepton flavour violation \cite{MEG,SINDRUM}. 
The cyan area represents the 
region of the parameter space which can be probed by 
the MEG experiment with the projected sensitivity 
to $\text{BR}(\mu \rightarrow e \gamma) = 10^{-13}$.
}
\label{fig:1}
\end{center}
\end{figure}
%%%%%%%%%%%%%%%%%%%%%%%%%%%%%%%%%%

%%%%%%%%%%%%%%%%%%%%%%%%%%%%%%%%%%%
\begin{figure}[t]
\begin{center}
\includegraphics[width=16cm,height=10cm]{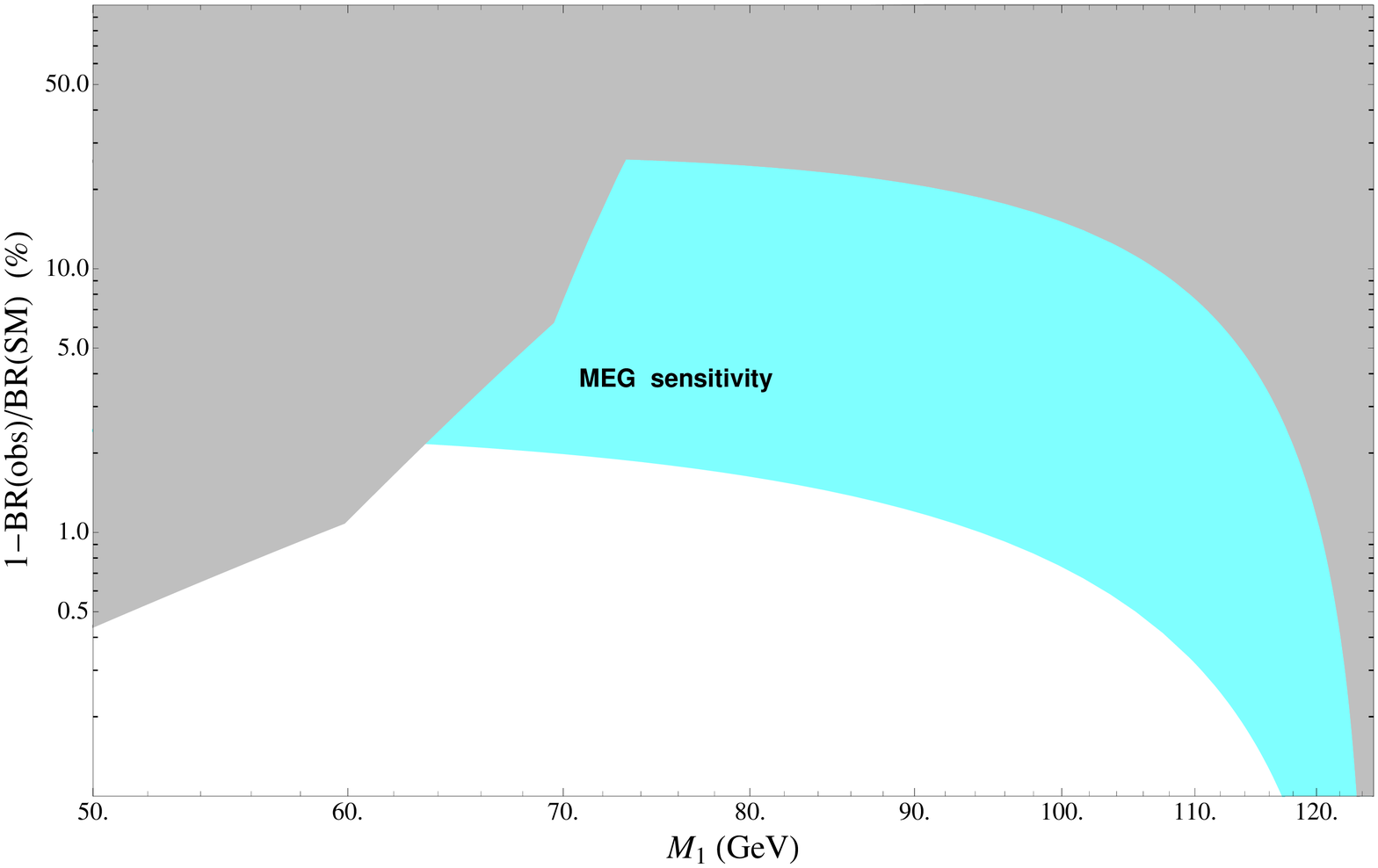}
\caption{Relative reduction of the Standard Model Higgs boson 
branching fraction to a generic channel for $m_{h}=125$ GeV. 
The color convention is the same 
as in Fig.~\ref{fig:1}.}
\label{fig:3a}
\end{center}
\end{figure}
%%%%%%%%%%%%%%%%%%%%%%%%%%%%%%%%%%

\section{New Higgs Decay Channels}

The decay rate of the Higgs boson to a SM fermion-antifermion pair 
is given by, at leading order in QCD corrections,
%%%%%%%%%%%%%%%%%%%%%%%%%%%
\begin{equation}
	\Gamma(h\to f \overline{f})\;=\; \frac{1}{16\pi}\,\left(\frac{m_{f}^{2}}{v^{2}} \right)\,m_{h}\,\left(1-\frac{4\, m_{f}^{2}}{m_{h}^{2}} \right)^{3/2}\,N_{c}(f)\,,
\end{equation}
%%%%%%%%%%%%%%%%%%%%%%%%%%%
%
with the usual color factor $N_{c}(f)$ equal to 1 and 3 
in the case of final state leptons and quarks, respectively. 
For a light Higgs particle, $m_h<160$ GeV, the dominant decay
channel is $h\rightarrow b\bar b$, which involves the Yukawa
coupling $y_{b}=m_{b}/v\simeq 0.024$.

 In the type I see-saw scenario of interest, 
the Higgs boson can also decay into a light and
a heavy pseudo-Dirac neutrino $N_{PD}$ 
provided $M_{1}<m_{h}$. 
In this case, the Higgs decay rate
is directly related to the neutrino Yukawa coupling $y$ 
defined in eq.~(\ref{ymax2}). 
Indeed, from the Lagrangian eq.~(\ref{NH}) and eq.~(\ref{ymax2}) we obtain:
%%%%%%%%%%%%%%%%%%%%%%%%%%%%%%%%%%
\begin{eqnarray}
	\Gamma(h\to\nu N) &\equiv&\sum_{\ell=e,\mu,\tau}\,  \left( \Gamma(h\to \nu_{\ell L}\, \overline{N_{PD}})+
	\Gamma(h\to \overline{\nu_{\ell L}}\,N_{PD}) \right) \nonumber\\
	& = & 
\frac{1}{16\pi}\, y^{2}\, m_{h}\,
\left( 1-\frac{M_{1}^{2}}{m_{h}^{2}} \right)^{2}\nonumber\,.
\end{eqnarray}
%%%%%%%%%%%%%%%%%%%%%%%%%%%%%%%%
%
Taking as benchmark values $m_{h}=125$ GeV and $M_{1}=100$ GeV, 
we obtain that  $\Gamma(h\to\nu N)/\Gamma(h\to b\overline{b}) 
\simeq 0.19~ (y/0.05)^{2}$. 
Hence, the decay channel $h\to \nu N$ 
could have a sizable branching
ratio if the upper limit on the Yukawa coupling $y$,  
obtained using the MEG upper bound on the  
$\mu\rightarrow e\gamma$ decay rate and
quoted in eq.~(\ref{yupIH}), is saturated. 
Conversely, the search for the Higgs decay $h\to \nu N$ can provide
limits on the parameters of the low scale see-saw model which are
competitive to those from the searches for the $\mu\to e\gamma$ decay, 
when  $m_h>M_1$. On the other hand, in the case $M_1>m_h$ 
the exotic Higgs decay channels are,
$h\to\nu N\to \nu\, \nu\, Z,\; \nu\, \ell\, W$ 
which have a rate suppressed by the fourth power of $y$ 
as well as by the three-body 
decay phase space. In view of the 
present upper limit on $y$
obtained from the existing experimental upper bounds on the rates of the 
lepton flavour violation processes 
$\mu \rightarrow e  \gamma$, $\mu \rightarrow 3e$ and 
$\mu - e$ conversion in nuclei  \cite{Ibarra:2011xn,Dinh:2012bp}, 
we conclude that the decay rates in these channels are too small 
to produce any observable effect. Hence, we will concentrate 
in what follows just on the possibility $M_1 < m_h$.

 We show in Fig.~\ref{fig:1} the values of $y$ as function 
of the see-saw scale $M_{1}$ corresponding to different 
values of $\text{BR}(h\rightarrow \nu N)$, for a fixed value 
of the Higgs boson mass, $m_{h}=125$ GeV. We also show the 
excluded region (grey area) by 
the results of 
i) the search for the $\mu\rightarrow e\gamma$ decay
with the MEG experiment \cite{MEG}, 
ii) the search for $\mu-e$ conversion in Ti \cite{SINDRUM}  
and iii) the search for heavy singlet neutrinos in 
$Z$ boson decays at LEP2~\cite{LEP2}.
It follows from the plot that the present limits 
on the low scale see-saw mechanism do not 
preclude the possibility of a Higgs boson 
decaying into a heavy and a light neutrino 
with a branching ratio which can be as large as 20\%, 
which, as we will see in the next section, can be observed 
at the LHC. Alternatively, the search for the exotic 
Higgs decay $h\rightarrow \nu N$ could provide 
the strongest limits on the parameter space of 
the low scale see-saw mechanism for RH masses 
smaller than the Higgs mass. We also show in the plot 
as a cyan area the projected sensitivity reach 
of the MEG experiment searching 
for the $\mu\to e\gamma$ decay 
with a branching ratio $\text{BR}(\mu\to e\gamma) \gtap 10^{-13}$,
which may allow to exclude 
$\text{BR}(h\to\nu N)\gtrsim 1\%$ for $M_{1}\gtrsim 100$ GeV.  

Furthermore, opening a new decay channel also 
modifies the branching ratios of the Higgs decay to a 
generic channel $X$ with respect to the 
corresponding SM prediction (BR(SM)):
%%%%%%%%%%%%%%%%%%%%%%%%%%%%%%%
\begin{equation}
	\text{BR}(h\to\nu N)\;\equiv\;\frac{\Gamma(h\to\nu N)}{\Gamma(h\to\nu N)+\Gamma_{\rm tot}^{\text{SM}}}\;=\;1-\frac{\rm BR(obs)}{\rm BR(SM)} \,\,\,\,\, ,
	\label{brhtoN}
\end{equation}
%%%%%%%%%%%%%%%%%%%%%%%%%%%%%
%
$\Gamma_{\rm tot}^{\text{SM}}$ being the total 
decay width of the Higgs boson in the Standard Model \cite{SMDecay}.
We show in Fig.~\ref{fig:3a} the maximal relative 
change of the branching fraction into a generic 
SM final state which arises in the low scale see-saw model, 
which is allowed by the current bound on the 
rate of the $\mu\to e\gamma$ decay. 
It follows from the plot that deviations as large as 
25\% are possible in this model.\footnote{ Notice that in this class of 
see-saw scenarios, in the case of IH light neutrino mass spectrum, 
a strong suppression of $\mu-e$ transitions might be possible 
for particular values of the CP violating phases in the neutrino 
mixing matrix if $0.15\lesssim\sin\theta_{13}\lesssim 0.2$ 
\cite{Dinh:2012bp}.  In this case, the best upper limit 
on the neutrino Yukawa coupling follows 
from the EW precision data:
$y\lesssim 0.06 M_{1}/(100~\text{GeV})$ \cite{Ibarra:2011xn}. 
This bound corresponds to $\text{BR}(h\to\nu N)\lesssim 34\%$ 
for $M_{1}\gtrsim 72$ GeV.} 
We also show in the plot the maximal relative change 
allowed if the MEG experiment reaches the sensitivity 
$\text{BR}(\mu\to e\gamma)\sim 10^{-13}$ 
without finding a positive signal. 
Conversely, the detection of a positive signal of 
$\mu\to e\gamma$ decay at MEG would imply the 
possibility of deviations from the SM branching 
ratios larger than 2\%, up to 25\%, 
for $70~\text{GeV}\lesssim M_{1}\lesssim 100~\text{GeV}$.

\mathversion{bold}
\section{Searches for the New Channel $h\to \nu N$ at LHC}
\mathversion{normal}

We have simulated with Madgraph~\cite{Alwall:2011uj} 
the process of production of a Higgs boson at the LHC, 
which decays $h\rightarrow \nu N$. 
We consider explicitly the final state 
with the heavy neutrino subsequently decaying 
into a charged lepton and an on-shell $W$ boson, 
which in turn decays producing two jets.\footnote{ The authors in \cite{Dev:2012zg} considered, within an inverse see-saw scenario, 
the alternative possibility to detect 
a heavy pseudo-Dirac singlet fermion 
through the fully leptonic decay mode:   
$h \to \overline{\nu_{\alpha L}}\,N_{PD}\,+
\,{\rm h.c.}\to \overline{\nu_{\alpha L}}\, 
\nu_{\beta L}\,\overline{\ell_{\gamma}}\, \ell_{\delta}\,+\,{\rm h.c.}$~.}
The processes of interest in our analysis are then: 
%%%%%%%%%%%%%%%%%%%%%
\begin{equation}
p\,p\;\to\; h\;\to\; \nu_{\alpha L}\, \ell_{\beta}^{+}\, j\, j\,, 
\bar{\nu}_{\alpha L}\, \ell_{\beta}^{-}\, j\, j\,,~\alpha,\beta=e,\mu,\tau\,.
\end{equation}
%%%%%%%%%%%%%%%%%%%%%% 
%
The branching fractions corresponding to the decays into 
the charged lepton $\ell_\alpha$ can be obtained 
from eqs. (\ref{mixing-vs-y}) and (\ref{mixing-vs-yIH}), 
the result being:
%%%%%%%%%%%%%%%%%%%%%
\begin{eqnarray}
\text{BR}(N_{PD} \to W\ell_\alpha)= \frac{m_{3}}{m_{2}+m_{3}}
    \left|U_{\alpha 3}+i\sqrt{m_{2}/m_{3}}U_{\alpha 2} \right|^{2} \sum_\beta \text{BR}(N_{PD} \to W \ell_\beta) \hspace{10pt}\text{for NH}\,,\label{BRNtoWN}\\
   \text{BR}(N_{PD} \to W\ell_\alpha)= \frac{m_{2}}{m_{1}+m_{2}}
    \left|U_{\alpha 2}+i\sqrt{m_{1}/m_{2}}U_{\alpha 1} \right|^{2} \sum_\beta \text{BR}(N_{PD} \to W \ell_\beta) \hspace{10pt} \text{for IH}\,.
\label{BRNtoWl}
\end{eqnarray}
%%%%%%%%%%%%%%%%%%%%%%%%%%%%%
%
In these equations (see, e.g., \cite{Atre:2009rg}), 
%%%%%%%%%%%%%%%%%%%%%%%%%%%%%%%
\begin{eqnarray}
\sum_\alpha \text{BR}(N_{PD} \to W\ell_\alpha) &=& \frac{(1-\mu_W)^2(1+2\mu_W)}{(1-\mu_W)^2(1+2\mu_W)+(1-\mu_Z)^2(1+2\mu_Z)}\,,\hspace{30pt}\text{if $\mu_Z<1$}\,,\nonumber\\
\sum_\alpha \text{BR}(N_{PD} \to W\ell_\alpha)&=& 1\,,\hspace{30pt}\text{if $\mu_Z>1$}\,,
\label{BrNtoW}
\end{eqnarray}
%%%%%%%%%%%%%%%%%%%%%%%%%%%%
%
where $\mu_W = (\frac{m_W}{M_1})^2$ and $\mu_Z = (\frac{m_Z}{M_1})^2$.

In our analysis we will only consider final states 
involving $e$ and $\mu$ due to the lesser 
efficiency in identifying $\tau$ leptons. Then,  
the total branching fraction of the process of interest is:
%%%%%%%%%%%%%%%%%%%%%%%%
\begin{eqnarray}
\text{BR}_\text{Total} &=& \text{BR}(h\to e^- \bar{\nu}jj)+\text{BR}(h\to \mu^- \bar{\nu}jj)+\text{BR}(h\to e^+ \nu jj)+\text{BR}(h\to \mu^+ \nu jj)\nonumber\\
               &=& \text{BR}(h\to \nu N) \,\left[~\text{BR}(N \to We)+\text{BR}(N \to W\mu)~\right]\,\text{BR}(W\to jj)\,,
\end{eqnarray}
%%%%%%%%%%%%%%%%%%%%%%%%%%%
%
which can be calculated from eqs.~(\ref{brhtoN}), 
(\ref{BRNtoWN}), (\ref{BRNtoWl}) and (\ref{BrNtoW}), 
with $\text{BR}(W\to jj)=0.676$ \cite{PDG10}. 
To estimate the relative branching ratio 
of the decay of the heavy neutrinos into $e+\mu$ flavours, 
we show in Fig.~\ref{BrtoWemu} the upper limit on the 
coupling $y$ for different values of the relative 
branching ratio, calculated using eqs.~(\ref{BRNtoWN}) and (\ref{BRNtoWl}) 
by taking the best fit values
of the neutrino oscillation parameters
\cite{Tortola:2012te} and 
varying the Dirac and Majorana phases of the 
neutrino mixing matrix  between 0 and $2\pi$.
We find that for the values of the Yukawa coupling 
that saturate the limits in eqs.~(\ref{yupNH}) and 
(\ref{yupIH}), the relative branching ratio into $e+\mu$ 
is approximately equal to $0.94$ for the IH and is in 
the range $0.20-0.80$ for NH. We will then 
use for our analysis the values 
$(\text{BR}(N \to We)+\text{BR}(N \to W\mu))/
\sum_\alpha \text{BR}(N \to W \ell_\alpha)=0.55$ and 0.94 
for NH and IH, respectively.
%%%%%%%%%%%%%%%%%%%%%%%%%
\begin{figure}[t]
	\centering
	\includegraphics[scale=1, width=10cm]{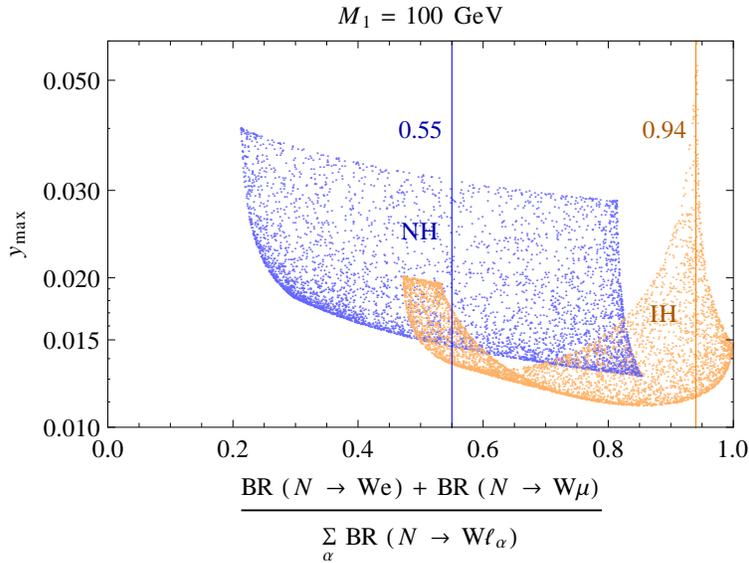} 
	\caption{Upper limit on the Yukawa coupling for various values of the relative branching fraction for decays into $e$ and $\mu$ for normal hierarchy (blue) and for inverted hierarchy (orange) and $M_1=100$ GeV. We also show in the plot the benchmark points taken in our analysis.
}   \label{BrtoWemu}
\end{figure}
%%%%%%%%%%%%%%%%%%%%%%%%

Now we define the signal identification and the 
corresponding reconstruction algorithm. 
Since our channel is one charged lepton, 
two jets plus missing energy, and following the 
detector coverage for the LHC experiments, 
we apply the following basic kinematical acceptance  
on the transverse momentum $p_{T}$, rapidity $\eta$
and the particle separation $\Delta R$:
%%%%%%%%%%%%%%%%%%%%%%%%%%%%%%%
\begin{eqnarray}
p_T(\ell) &>& 10 ~\text{GeV},  \hspace{10pt} |\eta_\ell |< 2.5\,,\nonumber\\
	\label{cutsLHCa}
	p_T(j) &>& 15 ~\text{GeV},  \hspace{10pt} |\eta_j |< 2.5\,, \\
	\Delta R(jj) &>& 0.4, ~ \Delta R(j\ell) > 0.4\,, \nonumber
\end{eqnarray} 
%%%%%%%%%%%%%%%%%%%%%%%%%%%
%
where the particle separation is defined as $\Delta R \equiv  \sqrt{(\Delta \phi)^2+(\Delta \eta)^2}$, $ \Delta \phi$ and $\Delta \eta$ being the azimuthal angular separation and the rapidity difference between two particles.  To further simulate the detector effects, we assume that the lepton and jet energies are smeared with a Gaussian distribution according to
%%%%%%%%%%%%%%%%%%%%%%%%%%
\begin{equation}
	\frac{\delta E}{E} = \frac{a}{\sqrt{E/\mbox{GeV}}}\oplus b\,,
	\label{smearing}
\end{equation}
%%%%%%%%%%%%%%%%%%%%%%%%%
%
\noindent
where $a_\ell = 5\%$ and  $b_\ell = 0.55\%$ for leptons, while 
$a_j = 100\%$ and $b_j = 5\%$ for jets \cite{Ball:2007zza}. 

 In order to construct efficient cuts 
to further reduce the background we have simulated the signal 
for $M_1=$ 90 GeV, 100 GeV, 110 GeV and 120 GeV 
using center of mass energies of 8 TeV and 14 TeV.
We show the corresponding normalized differential cross sections, after including the smearing, for the missing transverse energy ${\not} E_T$ (Fig. \ref{etdf}), for the total invariant mass $m_{jj\ell}$ (Fig. \ref{mjjldf}), which is peaked at the heavy neutrino mass, for the invariant mass of the jets $m_{jj}$ (Fig. \ref{mjjdf}), which is peaked at the $W$ boson mass, and for the reconstructed transverse mass $m_T$ (Fig. \ref{mtdf}), which has a Jacobian peak at the Higgs boson mass. From these distributions, it follows that the following cut on the missing transverse energy
%%%%%%%%%%%%%%%%%%%%%%%%%% 
\begin{eqnarray}
	{\not} E_T > 10 \text{ GeV}\,,
	\label{cutsLHCb}
\end{eqnarray} 
%%%%%%%%%%%%%%%%%%%%%%
%
and on the reconstructed masses
%%%%%%%%%%%%%%%%%%%%%%%
\begin{eqnarray}
	  80 \text{ GeV}<m_{jj\ell} <130 \text{ GeV}\nonumber\,,\\
	 m_W-10 \text{ GeV}<m_{jj} <m_W+10 \text{ GeV}\,,	\label{cutsLHCc}\\
	 110 \text{ GeV}<m_T <130 \text{ GeV}\nonumber\,.
\end{eqnarray}
%%%%%%%%%%%%%%%%%%%%%%%%%%%%%%%
will not reduce significantly the signal for a wide range of RH neutrino masses.

%%%%%%%%%%%%%%%%%%%%%%%%%%%%%%
\begin{figure}[t]
	\includegraphics[scale=1, width=8cm]{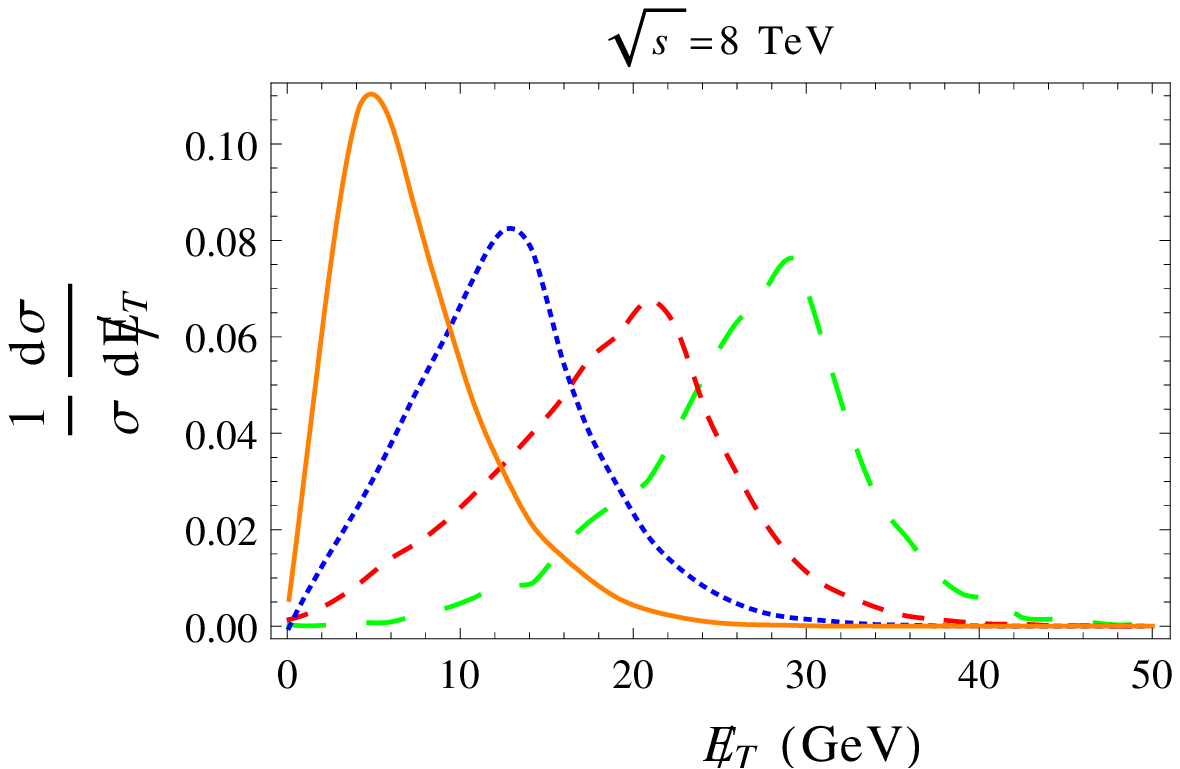} 
	\includegraphics[scale=1, width=8cm]{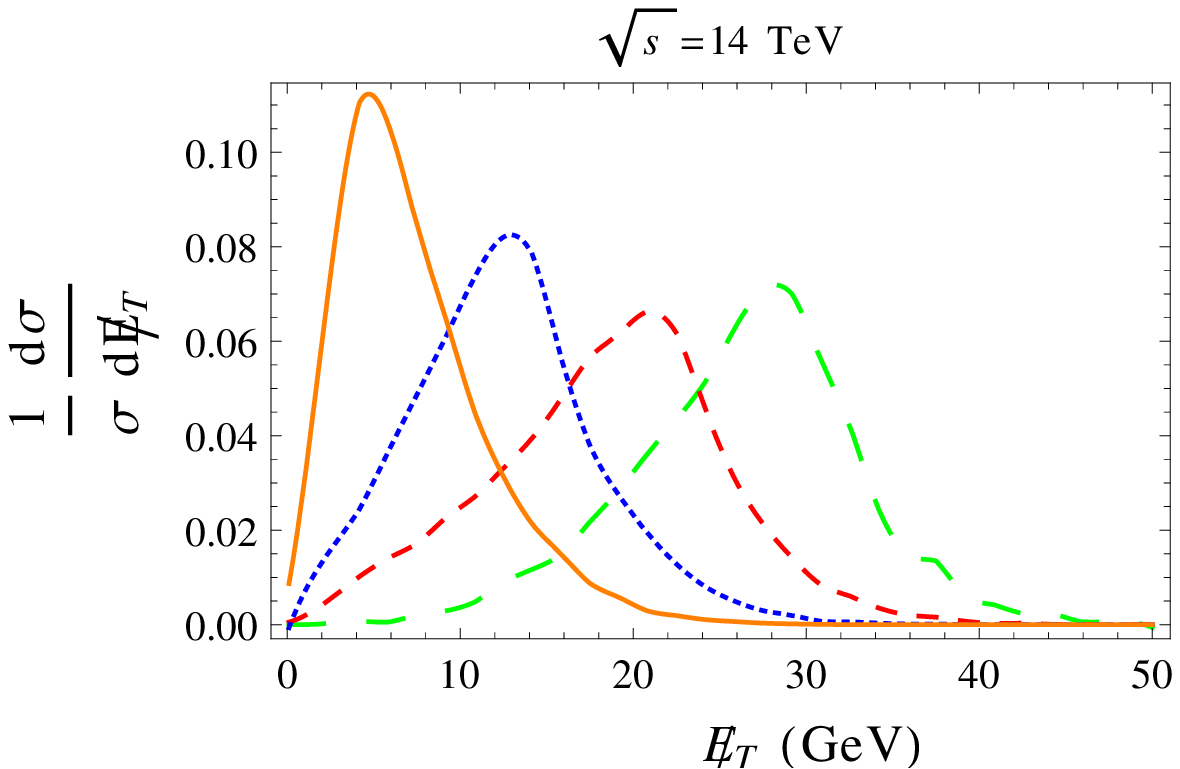} 
	\caption{Reconstructed normalized distributions $\frac{1}{\sigma} \frac{d\sigma}{d{\not} E_T} $ vs. the missing transverse energy, ${\not} E_T$, for various RH neutrino masses: $M_{1}=$120 GeV (continuous line), $M_1=$110 GeV (dotted line), $M_{1}=$100 GeV (short dashed line) and $M_{1}=$ 90 GeV (long dashed line).}
   \label{etdf}
\end{figure}
%%%%%%%%%%%%%%%%%%%%%%%%%%%%%%
%%%%%%%%%%%%%%%%%%%%%%%%%%%%%%
\begin{figure}[t]
	\includegraphics[scale=1, width=8cm]{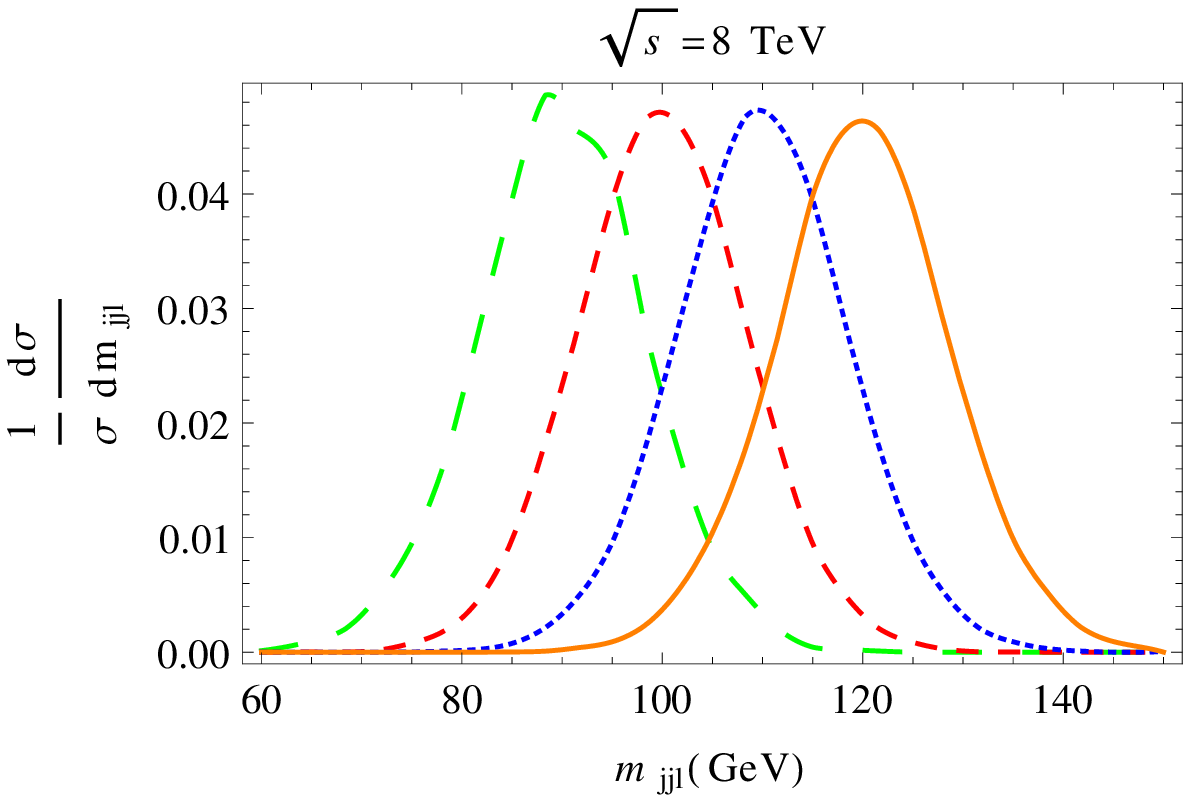} 
	\includegraphics[scale=1, width=8cm]{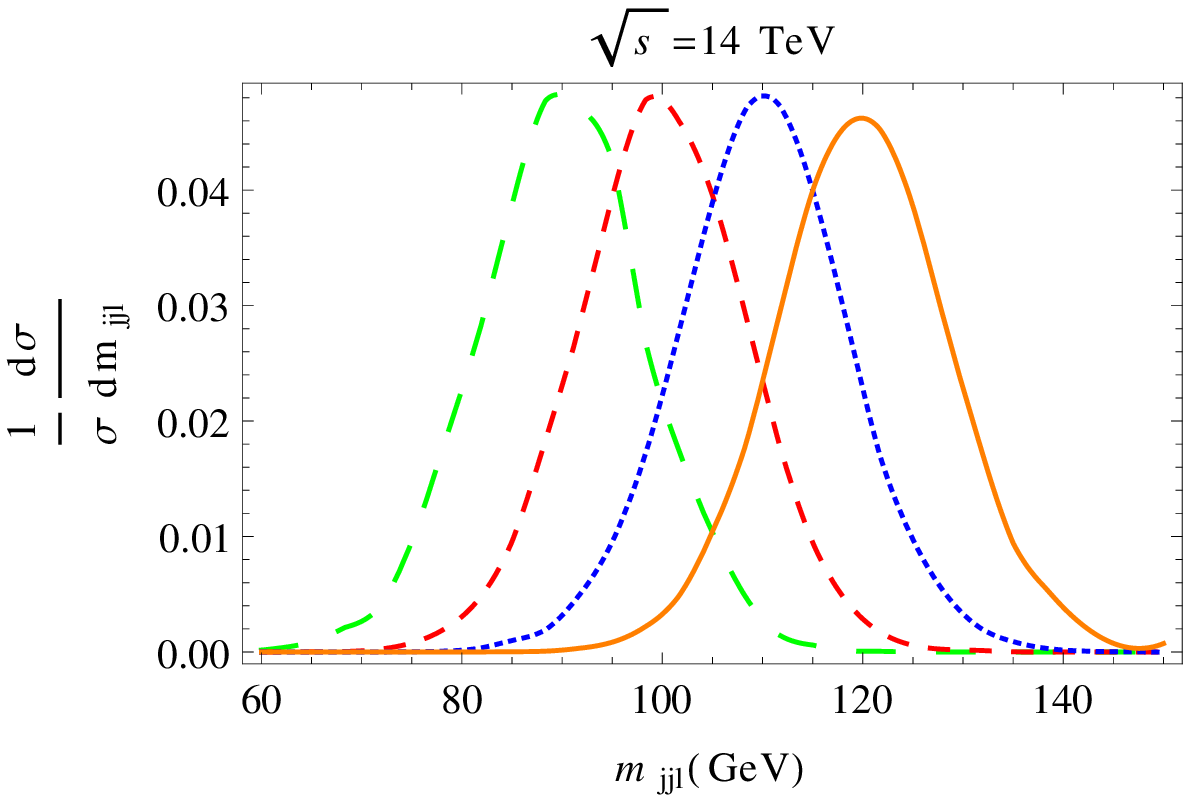} 
	\caption{Reconstructed normalized distributions $\frac{1}{\sigma}\frac{d\sigma}{dm_{jjl}}$ vs. the total invariant mass $m_{jjl}$. See Fig.~\ref{etdf} for details.}
   \label{mjjldf}
\end{figure}
%%%%%%%%%%%%%%%%%%%%%%%%%%%%%%%%
%%%%%%%%%%%%%%%%%%%%%%%%%%%%%%%%%
\begin{figure}[t]
	\includegraphics[scale=1, width=8cm]{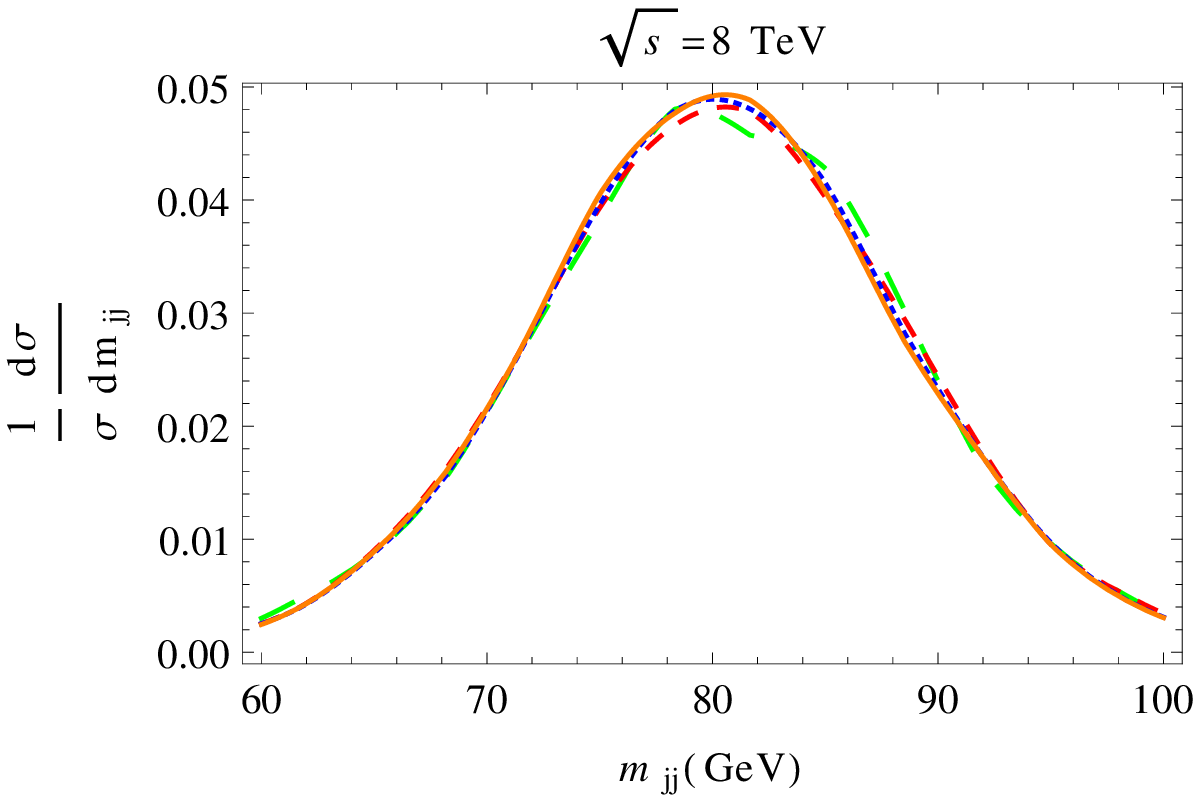} 
	\includegraphics[scale=1, width=8cm]{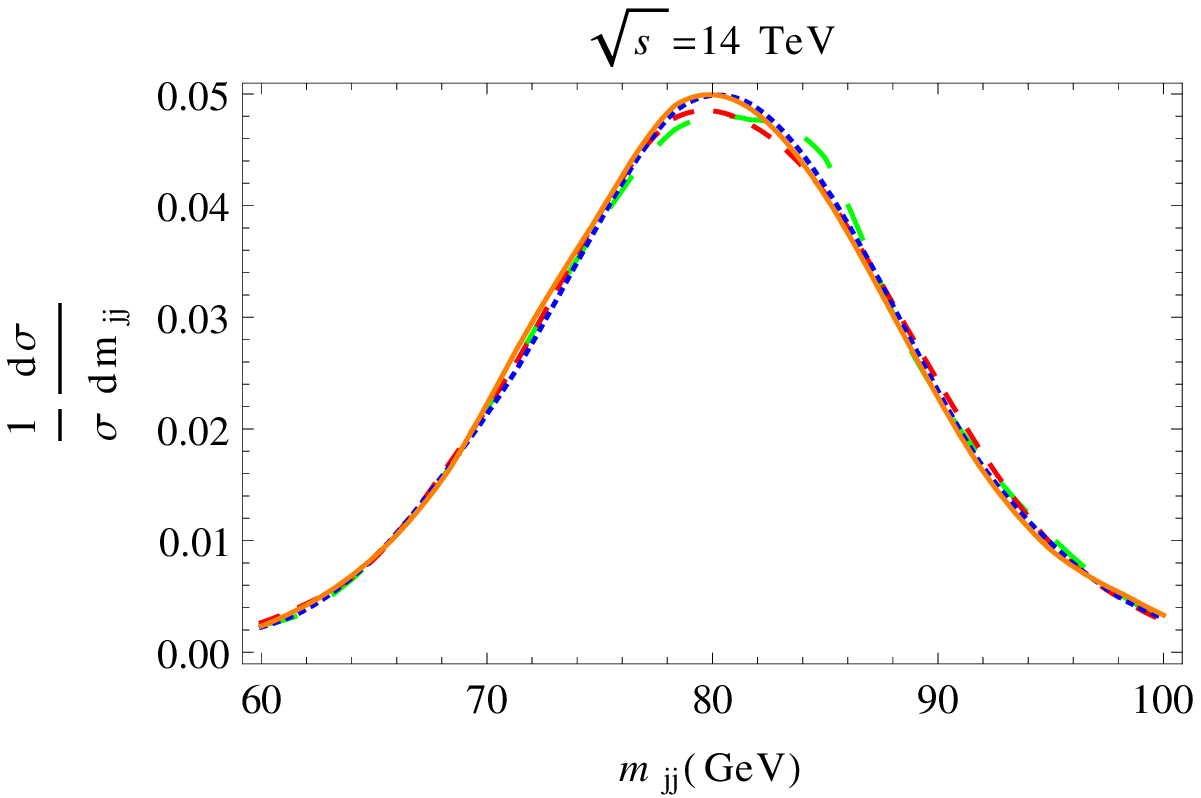} 
	\caption{Reconstructed normalized distributions $\frac{1}{\sigma}\frac{d\sigma}{dm_{jj}}$ vs. the invariant mass of the jets $m_{jj}$. See Fig.~\ref{etdf} for details.}
   \label{mjjdf}
\end{figure}
%%%%%%%%%%%%%%%%%%%%%%%%%%%%%%
\begin{figure}[t]
	\includegraphics[scale=1, width=8cm]{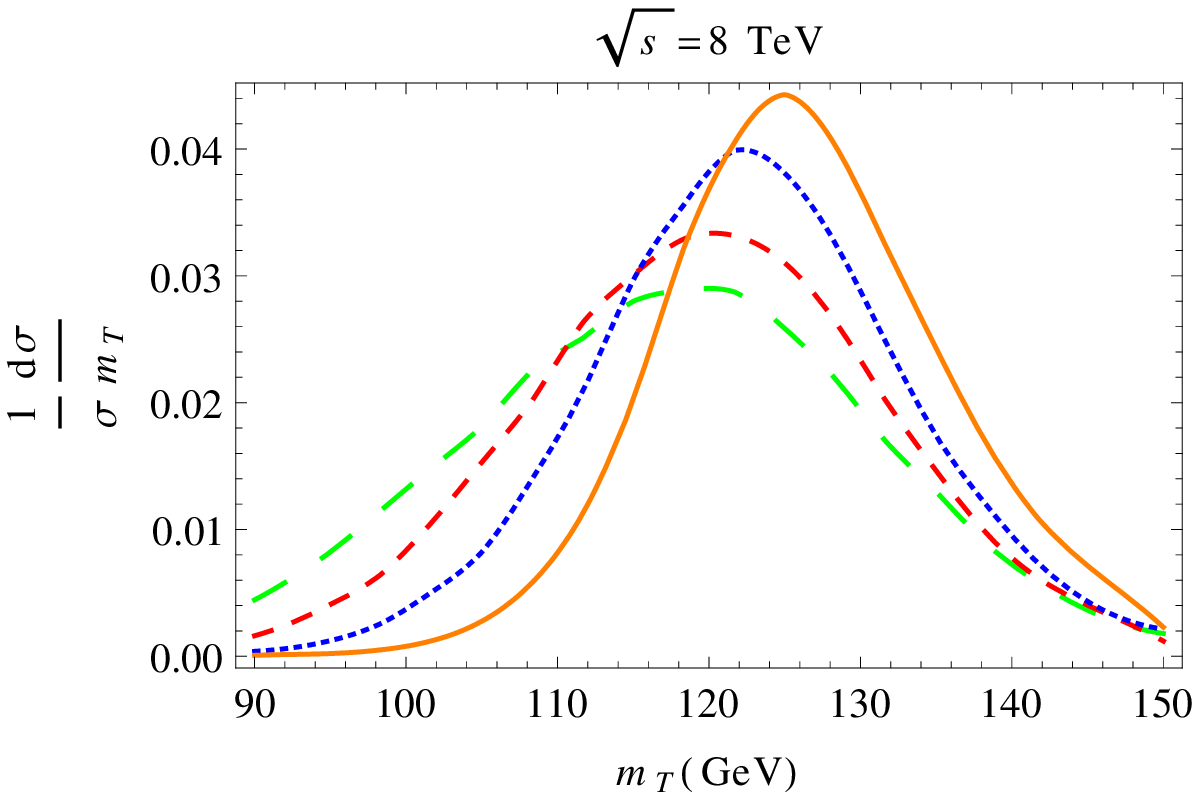} 
	\includegraphics[scale=1, width=8cm]{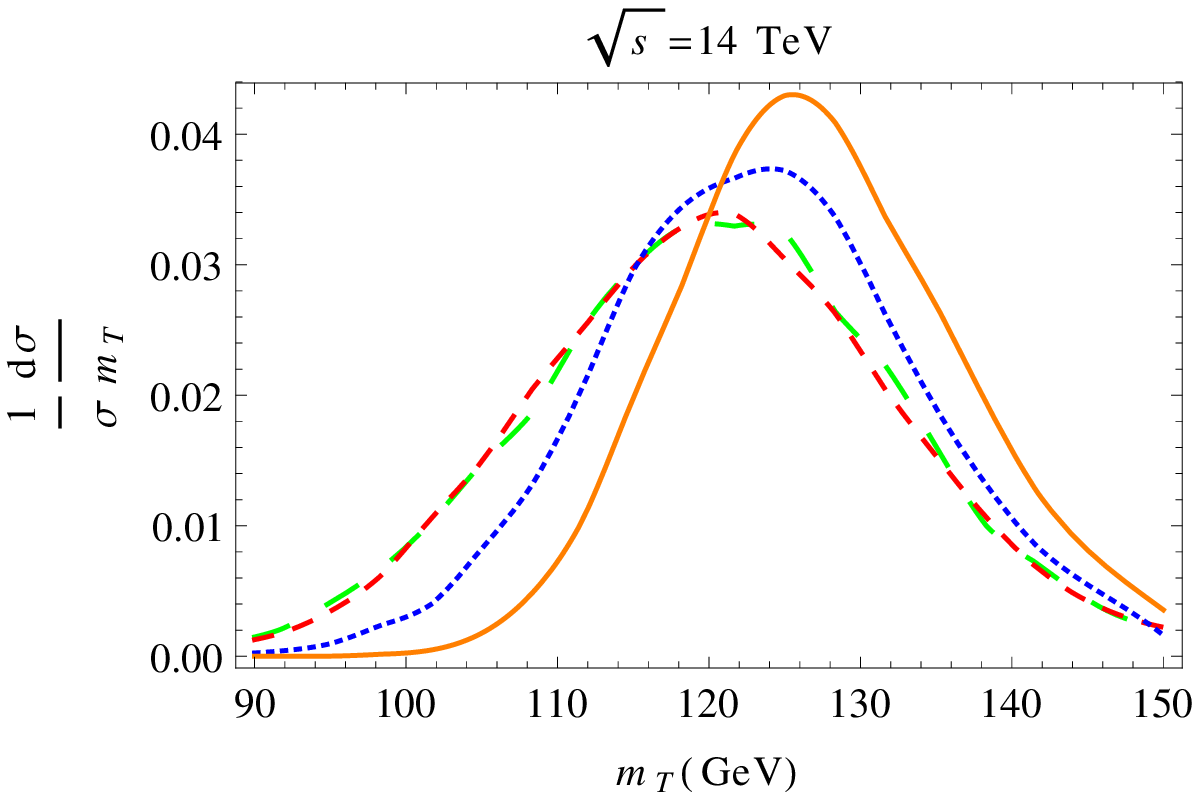} 
	\caption{Reconstructed normalized distributions $\frac{1}{\sigma}\frac{d\sigma}{dm_T}$ vs. the transverse mass $m_T$. See Fig.~\ref{etdf} for details.}
   \label{mtdf}
\end{figure}
%%%%%%%%%%%%%%%%%%%%%%%%

There are Standard Model backgrounds that lead to similar final states to our signal events, the most important being quark-gluon collisions when the final quark emits an off-shell $W$ boson which subsequently decays leptonically. Using Madgraph we have calculated the cross sections for the background processes. 
%We have calculated this and the other background processes using Madgraph. 
Here we ignore the faked leptons from heavy quarks like $b$ or $c$, assuming that our stringent separation requirement for the charged leptons will effectively remove them. 

%%%%%%%%%%%%%%%%%%%%%%%%%%%%%%%%%%
\begin{figure}[t]
	\includegraphics[scale=1, width=8cm]{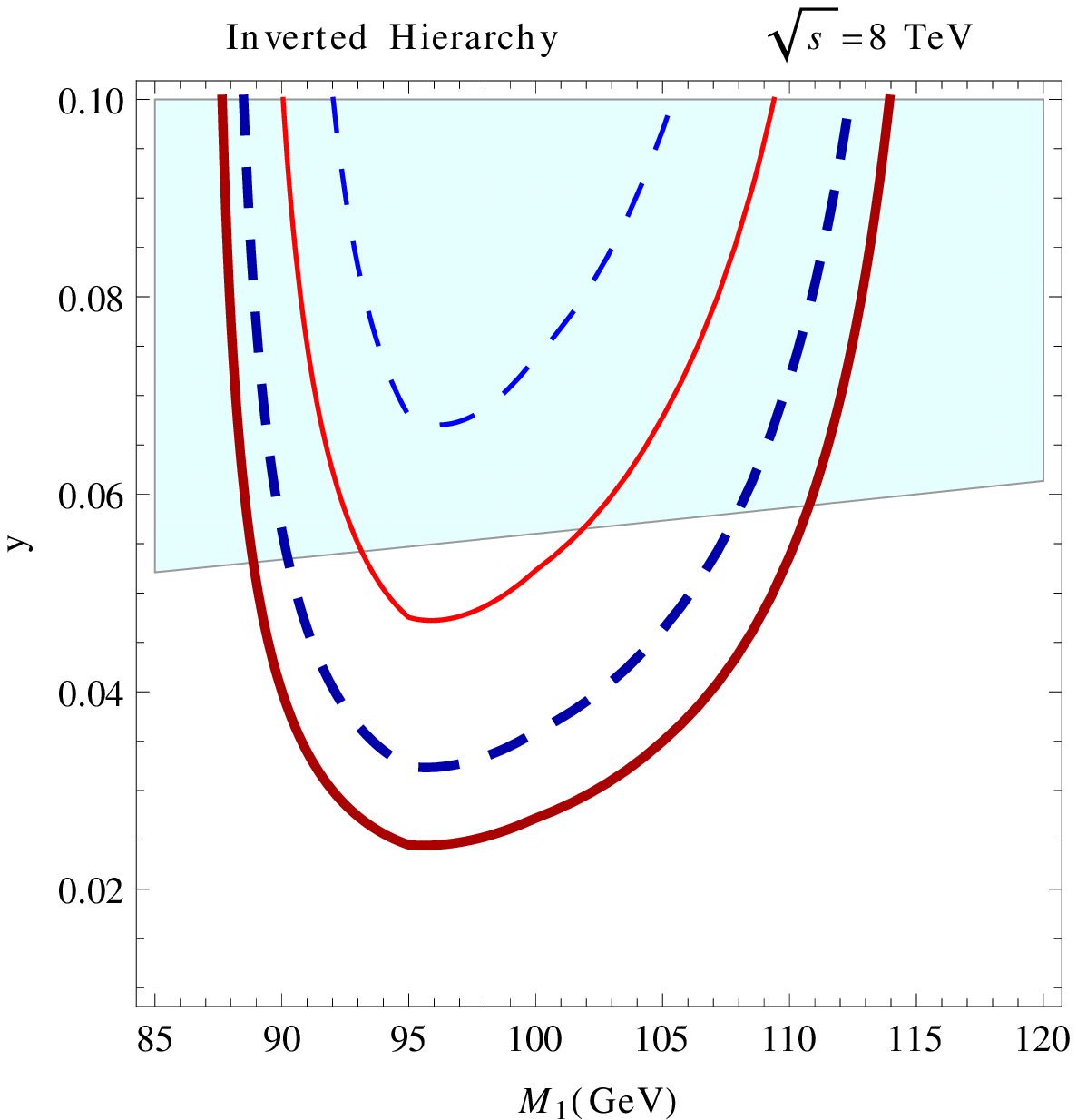} 
	\includegraphics[scale=1, width=8cm]{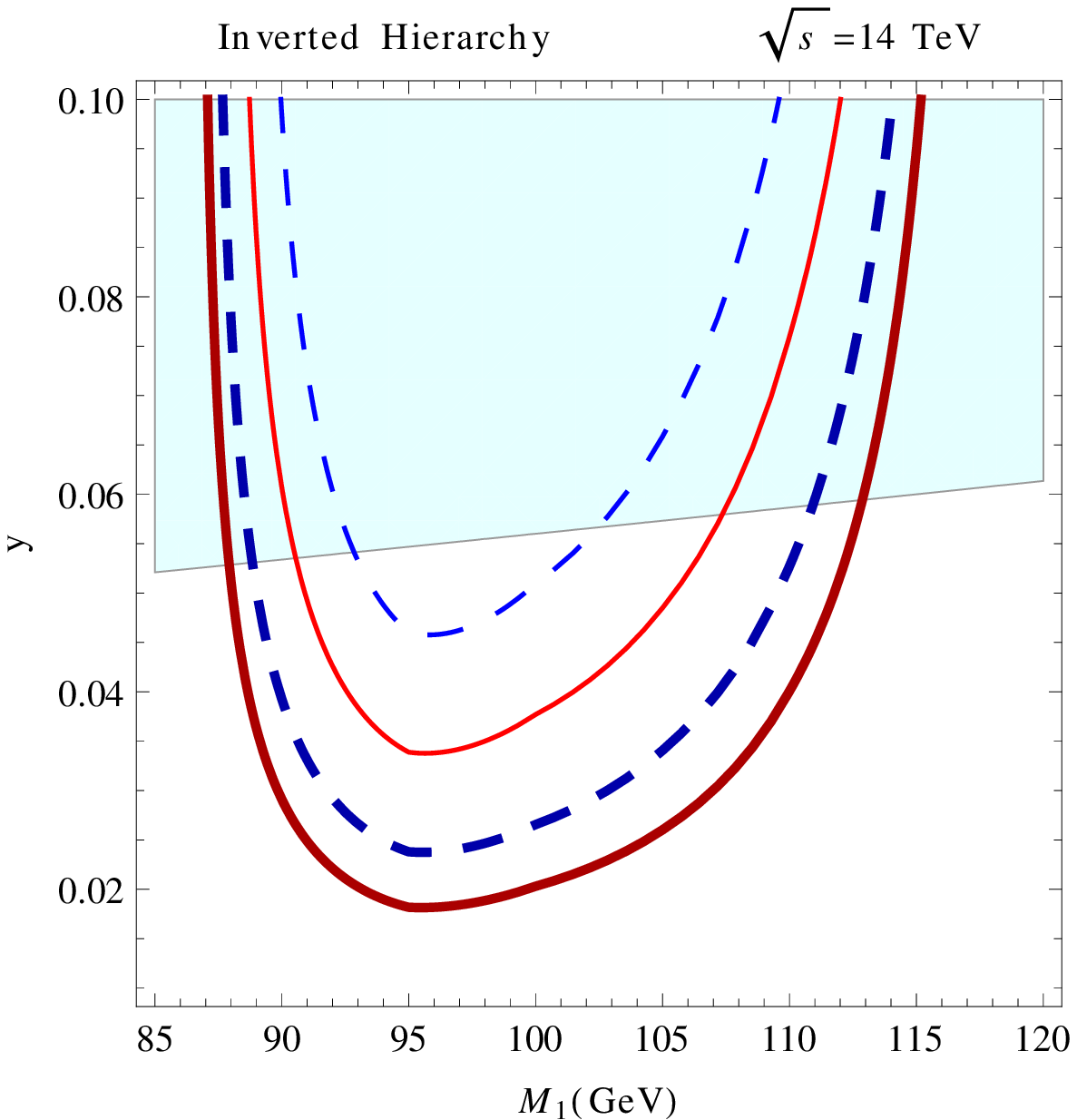} 
	\includegraphics[scale=1, width=8cm]{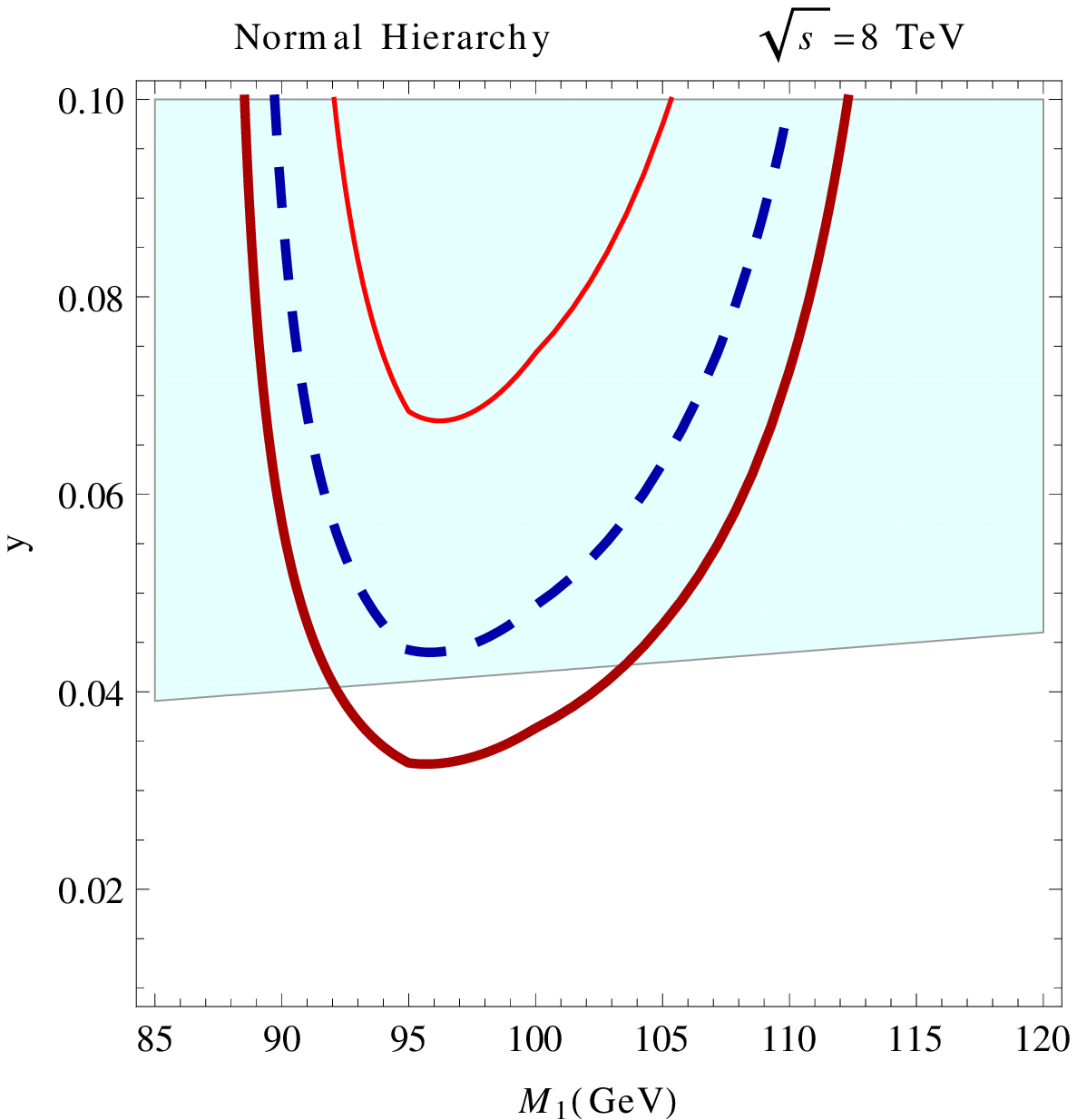} 
	\includegraphics[scale=1, width=8cm]{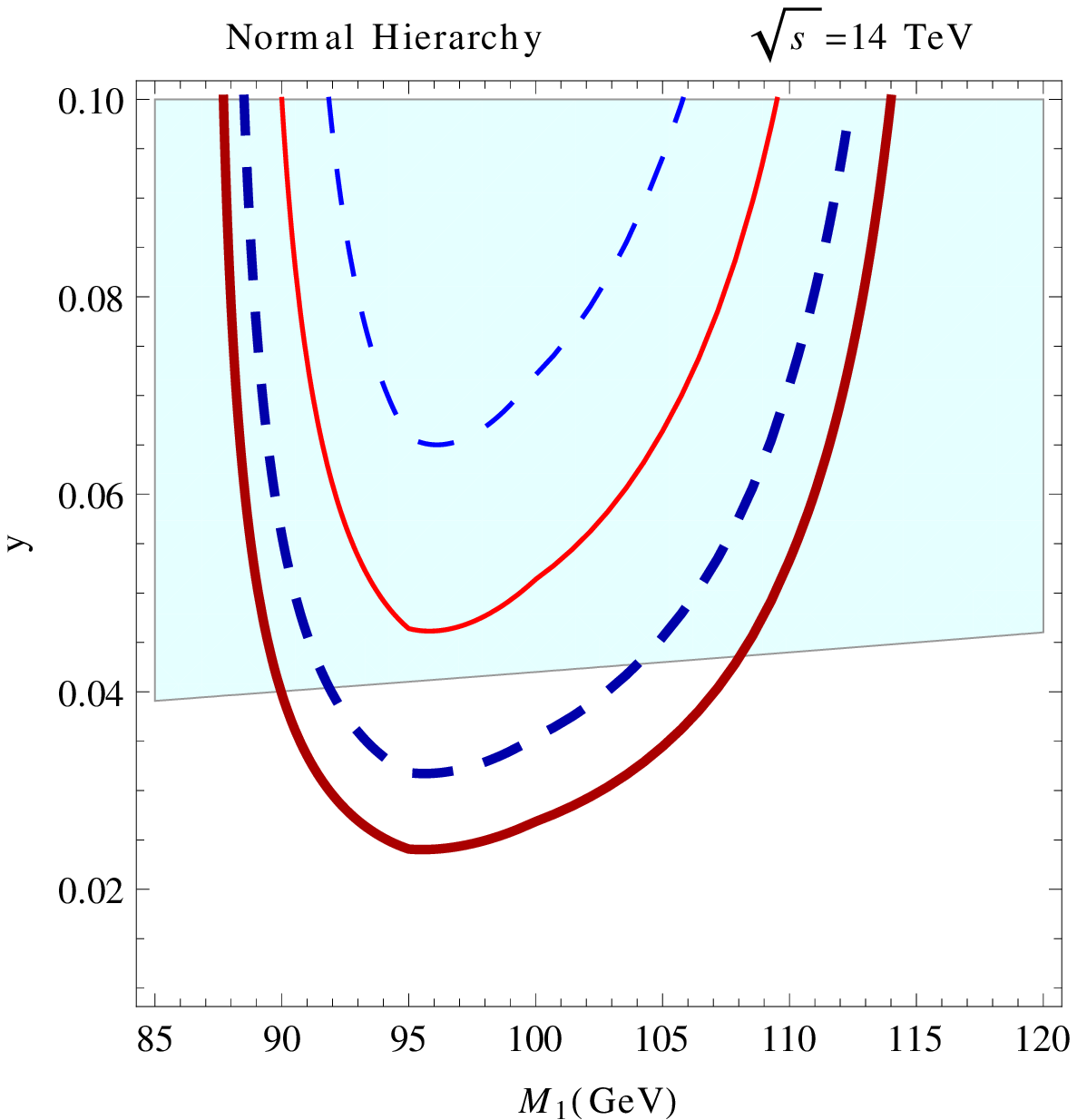} 
	\caption{Sensitivity of the LHC to the coupling $y$ vs $M_1$ at $3\sigma$ (continuous line) and $5\sigma$ (dashed line) and an integrated luminosity ${\cal L}=1~{\rm fb}^{-1}$ (thin line) and  ${\cal L}=10~{\rm fb}^{-1}$ (thick line), for inverted hierarchy (upper panels) and normal hierarchy (lower panels) and for $\sqrt{s}=8$ TeV (left panels) and $\sqrt{s}=14 $ TeV (right panels). The shaded region is excluded by the current experimental upper
	limit  $\text{BR}(\mu\rightarrow e\gamma)\leq 2.4\times 10^{-12}$  \cite{MEG}.
   \label{signi}}
\end{figure}
%%%%%%%%%%%%%%%%%%%%
%

For the sake of illustration, we  list
in table \ref{backvssig} the total cross sections 
for the background processes as well as for the signal 
(with masses $M_1$ of 90 GeV, 100 GeV and 110 GeV and $y=0.04$), 
after basic cuts, eq.~(\ref{cutsLHCa}), missing transverse energy cut, 
eq.~(\ref{cutsLHCb}), and mass cuts, eq.~(\ref{cutsLHCc}), 
for 8 TeV and 14 TeV. The reconstruction procedure outlined 
above effectively selects out the signal kinematics and 
substantially suppresses the SM backgrounds. 

We conservatively calculate the statistical significance to observe a signal
by
%%%%%%%%%%%%%%%%%%%%%%%%%%%%%%%%%%
\begin{equation}
	S = \frac{N_s}{\sqrt{N_s+N_b}}\, ,
\end{equation}
%%%%%%%%%%%%%%%%%%%%%%%%%%%
%
where $N$ corresponds to the number of events, and 
the subscripts $s$ and $b$ refer to the signal and 
the background respectively. Using the algorithm described
above, we have estimated the values of the neutrino 
Yukawa coupling $y$ that yield statistical significances 
of $3\sigma$ and $5\sigma$ for luminosities of 
$1\text{ fb}^{-1}$ and $10\text{ fb}^{-1}$ and center 
of mass energies of 8 TeV and 14 TeV. The results are 
shown in Fig. \ref{signi}. 

%%%%%%%%%%%%%%%%%%%%%%%%
\begin{table}[t]
	\centering
		\begin{tabular}{|c||c|c|c|c|} \hline
		\multicolumn{5}{|c|}{$\sigma (pb)$ } \\\hline 
 $\sqrt{s}$ 
& Process 
& \begin{tabular}{c} Basic cuts\\ Eq.~(\ref{cutsLHCa}) \\ NH (IH) \end{tabular} 
& \begin{tabular}{c} ${\not} E_T$ cut\\ Eq.~(\ref{cutsLHCb})\\ NH (IH) \end{tabular} 
& \begin{tabular}{c} Mass cuts\\ Eq.~(\ref{cutsLHCc}) \\ NH (IH) \end{tabular} 
\\\hline
\multirow{4}{*}{ 8 TeV }& Signal, $M_1= $ 90 GeV  & 0.061 (0.105) & 0.060 (0.103) & 0.033 (0.056) \\
                        & Signal, $M_1= $ 100 GeV & 0.132 (0.225) & 0.117 (0.200) & 0.067 (0.114) \\
                        & Signal, $M_1= $ 110 GeV & 0.051 (0.087) & 0.035 (0.059) & 0.019 (0.032) \\
                        & Background              & 1235 & 1189 &3.45 \\\hline
\multirow{4}{*}{ 14TeV }& Signal, $M_1= $ 90 GeV  & 0.155 (0.265) & 0.154 (0.263) & 0.083 (0.142) \\
                        & Signal, $M_1= $ 100 GeV & 0.339 (0.579) & 0.299 (0.511) & 0.171 (0.293) \\
                        & Signal, $M_1= $ 110 GeV & 0.130 (0.222) & 0.088 (0.151) & 0.048 (0.082) \\
                        & Background & 2635 & 2537 & 7.40 \\\hline
	\end{tabular}
	\caption{Effects of the kinematical cuts on the production cross section at the LHC for the signal $p\,p\to h\to j\,j\, \ell^+\,\nu$ + h.c.~and the corresponding SM background assuming normal (inverted) hierarchy. 
	We set the neutrino Yukawa coupling: $y=0.04$ .}
	\label{backvssig}
\end{table}
%%%%%%%%%%%%%%%%%%%%%%%%%%%%%%%%%
%

It follows from Fig.~\ref{signi} that  
the best sensitivity to the neutrino Yukawa coupling 
can be reached for a light neutrino mass spectrum  with inverted hierarchy. 
This is due to the fact that in this case 
${\rm BR}(N\to W e)+{\rm BR}(N\to W\mu)$ 
can be, as shown in Fig.~\ref{BrtoWemu}, plausibly close to one.
In particular, values of $y$ as small as 0.02 can be 
probed at LHC with a luminosity of $10\text{ fb}^{-1}$. 
Such values of the neutrino Yukawa coupling  can be 
directly tested    
by the MEG experiment \cite{MEG} searching 
for the $\mu\to e\gamma$ decay (see Fig.~\ref{fig:1}).
We find then an interesting interplay between collider searches
of RH neutrinos through Higgs decays and LFV observables, which may
be relevant for excluding type I see-saw scenarios 
with RH neutrino masses at the electroweak scale.

\section{Conclusions}

In this Letter we discussed quantitatively the 
possibility of producing and detecting 
at LHC  the heavy  
 $SU(2)_{L}\times U(1)_{Y}$ singlet 
fermions which appear 
in the context of TeV scale type I see-saw extension 
of the Standard Model with a mass $M$ at the electroweak 
scale. The recent discovery of a new 
scalar particle at LHC, which up 
to now  exhibits properties
that are consistent 
with those of the SM Higgs boson, 
opens the possibility of testing such kind 
of see-saw scenarios in collider
experiments through 
the observation of new exotic Higgs
decay channels in which the heavy fermions are produced. 

  The minimal version of the TeV scale 
type I see-saw scenario of interest contains 
two heavy Majorana neutrinos $N_{1,2}$ with masses $M_{1,2}$.
The requirement of reproducing the data on the neutrino 
masses and mixing determines the flavour structure 
of the neutrino Yukawa couplings 
as well as of the charged current and the neutral current 
weak interaction couplings of $N_{1,2}$ to the $W^\pm$ and $Z^0$ bosons
in the model. The existing low energy  phenomenological 
constraints on the indicated scenario can be satisfied if
the two heavy Majorana neutrinos form 
a pseudo-Dirac particle, $N_{PD} = 
(N_1 + iN_{2})/\sqrt{2}$, 
with $M_2 = M_1(1 + z)$, $z \ll 1$. 
As was shown in \cite{Ibarra:2011xn}, 
the type I see-saw scenario of interest is 
characterized by four real parameters: the mass $M_1\equiv M$, which  
sets the see-saw scale, the mass splitting parameter $z \ll 1$,  
a neutrino Yukawa coupling $y$ and a CP violation phase. 
Only two of these parameters - the mass $M$ and the Yukawa coupling $y$,
are relevant for the study performed in the present Letter.\footnote{The mass splitting $z$, for instance, 
is too small to have observables effects at LHC.
}

In this Letter we analyzed the 
 prospects of revealing the existence of
the additional SM singlet heavy 
Majorana neutrinos $N_{1,2}$, forming a  
pseudo-Dirac fermion $N_{PD} \equiv N$, 
in the case in which the Higgs particle 
is heavier than $N_{1,2}$ and decays 
with one charged lepton and two jets in the final state 
via the chain:
$h\to \nu\,N\to \nu\,\ell\,W\to \nu\,\ell\,jj$, 
where both $N$ and $W^{\pm}$ are on mass shell.
The results of our numerical analysis are
reported in  Table~\ref{backvssig}, where it is shown
that, after imposing the relevant
cuts on the total number of events,
the  QCD background can be drastically reduced
allowing the signal to be visible if enough luminosity
can be accumulated at the LHC.
The strength of the latter is strictly related 
to the values of the neutrino Yukawa coupling $y$ 
and the see-saw scale $M$.

We find that if $y\gtrsim 0.02$ and 
$90~\text{GeV}\lesssim M\lesssim 110~\text{GeV}$, then the heavy 
RH neutrinos (in the form of the pseudo-Dirac 
particle $N$) can be observed at LHC
with a statistical significance in 
the range of 3 to 5 $\sigma$ for a 
luminosity of 10 $\text{fb}^{-1}$ and a center 
of mass energy of 14 TeV (Fig.~\ref{signi}).
With a more sophisticated search strategy even 
smaller Yukawa couplings could be probed at the LHC.

 Sizable neutrino Yukawa couplings 
in the type I see-saw scenario considered
can also be probed by experiments searching 
for charge lepton flavour violation (LFV),
such as the MEG experiment which is devoted to the 
search for the $\mu\to e\gamma$ decay.
If the MEG experiment eventually observes the 
$\mu\to e\gamma$ decay with a branching ratio
$\text{BR}(\mu\to e\gamma) > 10^{-13}$,
the low scale type I see-saw scenario can
be directly tested at LHC through $h\to\nu\,N$ decays. 
Conversely, if no positive signal is detected 
in the MEG experiment and the upper limit 
$\text{BR}(\mu\to e\gamma) < 10^{-13}$ is obtained,
this will lead to a more stringent limit on the 
neutrino Yukawa coupling $y$ that will exclude 
the possibility of
producing and detecting the new heavy 
pseudo-Dirac neutrino $N$.

 As the results obtained in the present Letter show, 
the study of the properties of the Higgs boson observed at 
LHC will have important implications for the understanding 
of the origins of the neutrino masses and mixing as well.

\section*{Acknowledgments}
This work was supported in part by the INFN program on
``Astroparticle Physics'', by the Italian MIUR program on
``Neutrinos, Dark Matter and  Dark Energy in the Era of LHC''
and by the World Premier International Research Center 
Initiative (WPI Initiative), MEXT, Japan  (S.T.P.),
by the DFG cluster of excellence ``Origin and Structure of the Universe''
and the ERC Advanced Grant project ``FLAVOUR''(267104) (A.I.),
by the Graduiertenkolleg Particle Physics at 
the Energy Frontier of New Phenomena (C.G.C.)
and by the Funda\c{c}\~{a}o para a Ci\^{e}ncia e a
Tecnologia (FCT, Portugal) through the projects
PTDC/FIS/098188/2008,  CERN/FP/116328/2010
and CFTP-FCT Unit 777,
which are partially funded through POCTI (FEDER) (E.M.).

\end{document}